# PRODUCTION OF A LARGE DIAMETER ECR PLASMA WITH LOW ELECTRON TEMPERATURE

Mayuko Koga[*], Yasuhiro Hishikawa, Hayato Tsuchiya, Yoshinobu Kawai

*Interdisciplinary Graduate School of Engineering Sciences, Kyushu University,*

*6-1, Kasuga Koen, Kasuga, Fukuoka 816-8580, Japan*

ABSTRACT

A large diameter plasma over 300 mm in diameter is produced by electron cyclotron resonance (ECR) discharges using a newly developed large diameter chamber of 400 mm in inner diameter. It is found that the electron temperature decreases as nitrogen gas is added to pure argon plasma, which is considered that nitrogen molecules absorb the electron energy into the vibrational energy and lead to the decrease of the electron temperature. Moreover, it is found that the plasma uniformity is improved by adding nitrogen gas, suggesting that the adjunction of nitrogen gas is effective for producing uniform and low electron temperature plasma.

1. Introduction

Recently, a large diameter plasma with the diameter more than 300 mm has been required in industry. Electron cyclotron resonance (ECR) plasma [1-3] has great advantages for plasma application such as high electron density, low gas pressure operation and low contaminations compared with other plasma sources. However, the electron temperature in ECR plasma is relatively high so that the charge up damage is often caused on substrates. Thus, to produce a large diameter ECR plasma with low electron temperature is one of the most important subjects for plasma application.

In order to control the electron temperature, some methods have been reported such as grid method [4], pulse modulation [5] and use of low frequency microwaves [6]. Itagaki et al. succeeded in decreasing the electron temperature by using the mirror magnetic field and found that the decreasing effect was enhanced by adding nitrogen gas to argon plasma [7]. It is considered that nitrogen molecules absorb the energy of electrons and cause the low electron temperature [8]. However, the relationship between the electron temperature and the nitrogen gas concentration has not been investigated yet. Here, we report the production of a large diameter ECR plasma by using a newly developed large diameter chamber and the investigation of the relationship between the electron temperature and the nitrogen gas concentration.

2. Experimental setup

Figure 1 shows the schematic diagram of the experimental apparatus. The newly developed large diameter chamber was a cylindrical vacuum chamber made of stainless steel with an inner diameter 400.4 mm and a length 1200 mm. The magnetic coil assembly consisted of eight electromagnetic coils. Positions and currents of the coils were variable to form various magnetic field configurations. The frequency of microwaves was 2.45 GHz and the power could be increased up to 1.3 kW. Microwaves were launched into a chamber as a circular $TE_{11}$ mode through the tapered waveguide and the quartz window. The matching between the microwave circuit and the plasma was adjusted with the three-stub tuner in such a way that the reflected microwave power monitored by power monitor was as low as possible. Rotary pump and 2000 l/s turbo molecular pump were connected to evacuate the chamber. After evacuating to a base pressure less than $4\times10^{-6}$ Torr, argon and nitrogen gases were introduced into the chamber through mass flow controllers. The density and the temperature of electrons were measured with a cylindrical single Langmuir probe (1 mm diam. and 1 mm length).

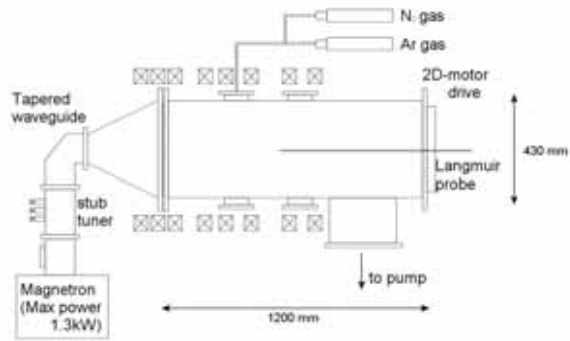

Figure 1 Schematic diagram of the experimental apparatus.

3. Results and discussion

Firstly, we attempted to generate a large diameter plasma by using a new device. In general, the ion saturation current is proportional to the electron density of plasma. Therefore, it is convenience way to measure the profile of the ion saturation current in order to know that of the electron density. Figure 2 shows the radial profile of the ion saturation current as a function of the incident microwave power. The gas pressure was kept at 2 mTorr. It is clearly seen that the large area plasma with diameter over 300 mm is obtained successfully. It is found that the radial profile of the ion saturation current changes from flat profile to convex profile as the microwave power is increased. This change is considered to be the effect of extra ordinary wave (X wave). It is known that X wave which propagates perpendicular to the magnetic field causes upper hybrid resonance and contributes to plasma generation near the wall region when the electron density becomes around the L cutoff, $1-2\times10^{11}$ cm$^{-3}$ [9]. In fact, the measured electron density where the radial profile changed was corresponding with this value.

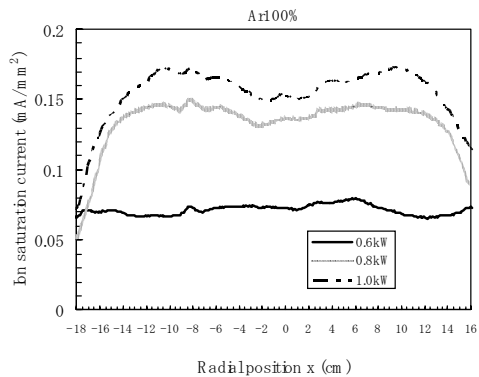

Figure 2 The radial profile of the ion saturation current as a function of the incident microwave power.

Ueda et al. found that the plasma uniformity is controlled by changing the gradient of the magnetic field at the resonance position [10]. This is because the power absorption is changed by changing the gradient of the magnetic field. Thus, in order to improve the plasma uniformity, the radial profiles of the electron density under different magnetic field configurations were investigated. Figure 3 shows radial profiles of the ion saturation current for different magnetic field configurations, where the gas pressure is 2 mTorr and the incident microwave power is 1.2 kW. As shown in Fig. 3, the radial profile of the plasma density changes from peaked profile (b) to flat profile (a) by changing the magnetic field configuration. In the case of magnetic field configuration (a), flat density profile with the uniformity about 15 % was obtained.

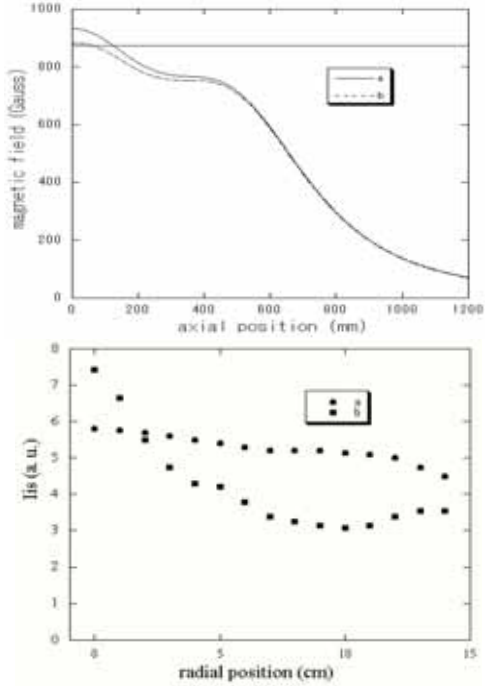

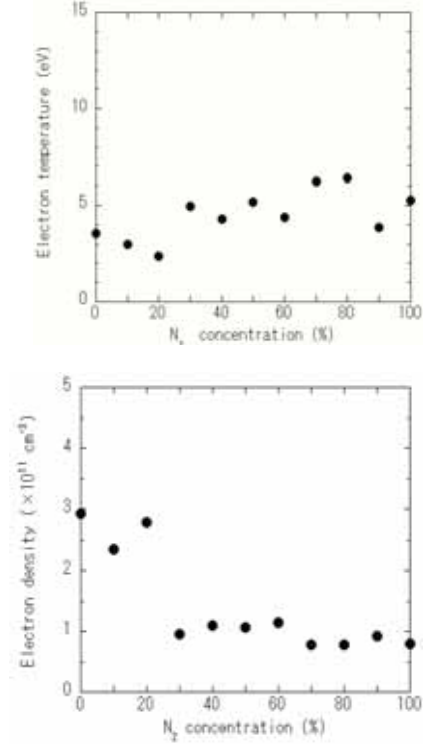

Figure 3 The radial profile of the ion saturation current (below) corresponding to the different magnetic field configuration (upper).

Figure 4 The nitrogen gas concentration dependences of the electron temperature (upper) and the electron density (below).

Secondly, we investigated the relationship between the electron temperature and the nitrogen gas concentration. Figure 4 shows the electron temperature and the electron density as the function of the nitrogen gas concentration. The nitrogen gas concentration $\alpha_{N2}$ was determined as follows; $\alpha_{N2}=S_{N2}/(S_{N2}+S_{Ar})$. Here $S_{N2}$ and $S_{Ar}$ denote the gas flow rate of nitrogen gas and argon gas, respectively. The total gas pressure was kept at 2 mTorr and the incident microwave power was 1.0 kW. It is found that the electron temperature decreases as the nitrogen gas concentration is increased up to 20 %. It is considered that the decrease of the electron temperature was due to the absorption of the electron energy into the vibrational energy of nitrogen molecules. As shown in Fig. 5, nitrogen molecules have a large vibrational cross section at low energy of 2 eV [11]. On the other hand, in the case of argon gas, the electron energy is mainly absorbed into the collision energy between electrons and argon atoms, which shows a large peak at high electron energy range of 10-100 eV. In our system, the electron temperature was below 10 eV. Therefore, it is considered that the effect of energy absorption by nitrogen molecules was larger than that by argon atoms and caused the decrease of the electron temperature. As shown in Fig. 4, the electron temperature begins to increase when the nitrogen concentration is increased more than 20 %. This increase of the electron temperature is considered as the result of the increase of the plasma loss, since nitrogen molecules and nitrogen atoms are light compared with argon atoms. It is well known that the electron temperature generally increases to sustain the plasma when the electron density decreases. In fact, as shown in Fig. 4, the electron density decreases with the increase of the nitrogen gas concentration. The other candidate is upper hybrid resonance excited by X wave. If the hybrid resonance occurs, electrons gain the energy from the wave and the electron temperature increases. Further measurement is needed for verification.

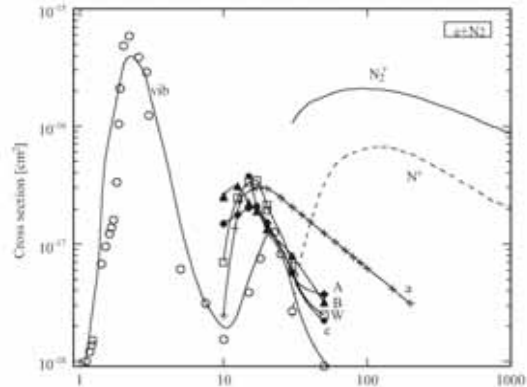

Figure 5 Cross sections for electron collisions with $N_2$: $N_2^+$ and $N^+$ are ionization; vib is vibrational excitation; A, B, W, c and a are excitation for the relevant states (reference 9).

Figures 6 and 7 show the pressure dependence of the electron temperature and the electron density, respectively. The incident microwave power was 1 kW. As shown in Fig. 6, the electron temperature in argon/nitrogen plasma is about 0.3 eV lower than that in pure argon plasma under all gas pressure conditions. Therefore, it is concluded that the addition of nitrogen gas for decreasing of the electron temperature is applicable in wide range of pressure. As shown in Fig. 7, the pressure dependence of the electron density in argon/nitrogen plasma is small compared with that in pure argon plasma. This result may be due to the large plasma loss of argon/nitrogen plasma. However, the electron density of argon/nitrogen plasma ($\alpha_{N2}$=10 %) is sufficient for plasma application (about $2 \times 10^{11}$ cm$^{-3}$).

Moreover, it was found that the radial uniformity of plasma was improved by adding nitrogen gas. Figure 8 shows the radial profile of the ion saturation current in the case of argon/nitrogen plasma. It is clearly seen that the radial profile became flatter compared with the case of pure argon plasma, Fig.2. The plasma uniformity in 300 mm diameter was improved from ±14.0 % to ±7.7 % by adding the nitrogen gas. Therefore, it is considered that the addition of the nitrogen gas is effective to generate the uniform plasma.

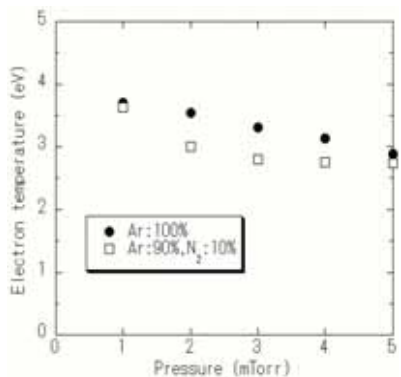 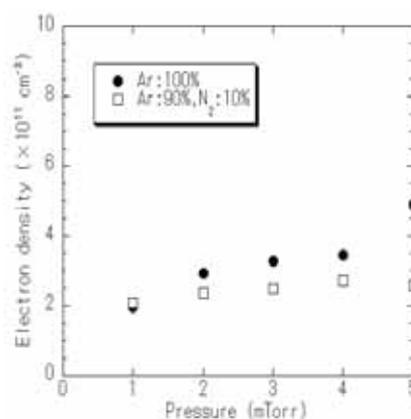

Figure 6 The pressure dependence of the electron temperature.

Figure 7 The pressure dependence of the electron density.

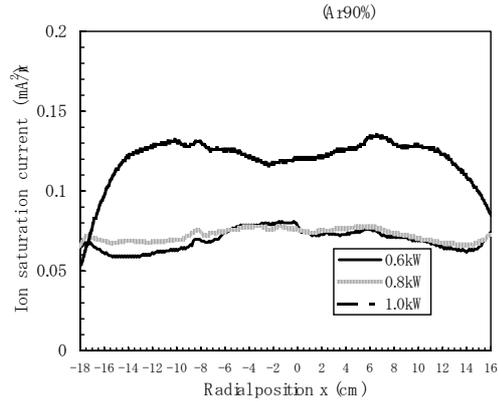

Figure 8 The radial profile of the ion saturation current as a function of the incident microwave power. (argon/nitrogen plasma).

4. Summary

In this paper, a large diameter ECR plasma was produced using a newly developed large diameter device and the relationship between the electron temperature and the nitrogen gas concentration was investigated, in detail. We succeeded in producing large diameter plasma with diameter over 300 mm. It was found that the plasma profile changed from flat to concave when the incident microwave power was increased. This was considered as the effect of X wave. It was also found that the plasma uniformity is improved by adjusting the magnetic field configuration.

On the other hand, the electron temperature decreased as the nitrogen gas concentration was increased up to 20 %. It was considered that the electron energy was absorbed into the nitrogen molecular vibrational energy and, as the result, the electron temperature decreased. It was found that the addition of nitrogen gas is applicable in wide range of the operating gas pressure and the electron density of argon/nitrogen plasma ($\alpha_{N2}$=10 %) is sufficient for plasma application. Moreover, it was found that the plasma uniformity is improved by adding nitrogen gas. Therefore, it is concluded that the addition of nitrogen gas to pure argon plasma is effective for producing a uniform plasma with low electron temperature.